\documentstyle[spie,epsf,12pt]{article}
\title{\hphantom{AAAAAAAAAAAAAAAAAAAAAAAAAAAAAAAAAAAAAAAAAAAAAAAAAAAAA}
Measures of Decoherence}
\author{{\Large Leonid Fedichkin, Arkady Fedorov and Vladimir Privman}}

\begin{document}

\maketitle

\centerline{Center for Quantum Device Technology,}

\centerline{Department of Physics and}

\centerline{Department of Electrical and Computer Engineering,}

\centerline{Clarkson University, Potsdam, New York\ 13699--5720}

\begin{abstract}
Methods for quantifying environmentally induced decoherence
in quantum systems are investigated. We formulate criteria for measuring
the degree of decoherence and consider several representative examples,
including a spin interacting with the modes of a
bosonic, e.g., phonon, bath. We formulate an approach based on the operator 
norm measuring the deviation of the actual density matrix from the ideal
one which would describe the system without environmental interactions.


\end{abstract}

\noindent{}{\bf Citation:\ } \ Proc. SPIE \underline{5105}, 243-254 (2003)

\noindent{}{\bf E-print:\ } \ \,\ cond-mat/0303158 at www.arxiv.org

\section{Introduction}
Evolution of a quantum system exposed to an environment is described by the
density matrix and deviates from the ideal, usually pure-state, dynamics.
In this work, we consider the problem of quantifying this deviation by
single numerical measures, derived from the full set of the density matrix
elements. Depending on their strength and nature, environmental interactions
can lead to various relaxation and even measurement-type processes. We
will use the term ``decoherence'' generically. Our focus will be on the
environmental effects that represent ``noise'' and cause small deviations
from the desired isolated-system dynamics. Establishment of the threshold
criteria for fault-tolerant quantum computation,\cite{Aharonov} has made 
estimation of the noise due to the environmental effects important for 
evaluating quantum computing systems.

Superposition of quantum states is crucial in utilizing quantum
parallelism for quantum computation. Therefore quantum algorithms usually deal with pure
or nearly pure states. There are actually two possible measures of decoherence:
deviation from {\em a\/} pure state, and deviation from a {\em particular\/} pure
state, or nearly-pure mixed state, which can be time-dependent. Even if the state of
the system remains pure, it might
deviate from the dynamics desired for a particular controlled quantum process.
The ultimate goal of the studies of the type reported here would be to
identify simple numerical measures of degree of decoherence occurring during the ``clock''
times of quantum computing gate functions, to compare
with the fault-tolerance requirements in various quantum error correction schemes.

Let us assume that a quantum system of interest, denoted by $S$, is
prepared at time $t=0$ in some pure quantum state $|\varphi\rangle$.
At times $t>0$, the system can be subject to noise, i.e., coupling
to the modes of the environment, but also to ``controlling'' interactions
required for quantum computing. Here we consider a system only exposed to the
environmental noise, because the degree of decoherence can be
usually approximately evaluated for time scales shorter than or comparable to the times
of quantum control. The quantum system can then be described by the reduced
density operator, $\rho (t)$, which can be obtained from the overall density operator
by tracing over all the environmental degrees of freedom.

For the initial
state, the density matrix has the form $  \rho (0) =
| \varphi  \rangle \langle\varphi  |$. The interaction with the
surrounding, which is usually assumed to be a large macroscopic
system in thermal equilibrium at temperature $T$, leads to thermalization of the
quantum system. The reduced density matrix of the system for large times
should approach
\begin{equation}
\rho \to {{e^{ - \beta   H_S } }}\Big / {{{\rm Tr}_{\, S} \left(e^{ -
\beta H_S } \right)}}, \qquad {\rm as} \;\; t \to \infty,
\end{equation}\par\noindent
where  $\beta = 1/kT$. Here $  H_S$  is the Hamiltonian of the system. In the energy 
eigenbasis, $|\phi_n \rangle$, the density matrix elements behave as follows,
\begin{equation}
\label{eq2}
\rho _{mn}  = \langle \phi_m | \rho | \phi_n \rangle \to 0, \qquad {\rm as} \;\; t \to \infty \qquad (m \ne n),
\end{equation}\par\noindent
\begin{equation}
\label{eq3}
\rho _{nn}  = \langle \phi_n | \rho | \phi_n \rangle \to e^{ -
\beta E_n }\Big / \sum\limits_k e^{ -
\beta E_k }, \qquad {\rm as} \;\; t \to \infty .
\end{equation}\par\noindent
Usually, thermalization sets in as the slowest relaxation process,
and its description requires a phenomenological Markovian assumption.\cite{therm4,therm2,therm1}

In this work, we consider several approaches to measuring the degree
of decoherence due to interactions with the environment. In Section 2, we
discuss the approach based on the asymptotic relaxation time scales. The entropy and
idempotency-defect measures are addressed in Section 3. The fidelity measure of
decoherence is discussed in Section 4. Next, in Section 5, we
present our results on the operator norm measures of decoherence.
Section 6 presents a discussion of an approach to eliminate the initial-state
dependence of the decoherence measures, as well as of extensivity properties 
for multiqubit systems.

\section{Relaxation Time Scales}
Markovian approximation schemes typically yield exponential
approach to the limiting values of the density matrix elements for
large times.\cite{therm4,therm2,therm1} For the two-state
system, this defines the thermalization time scales $T_1$ and
$T_2$, associated, respectively, with the approach by the diagonal
(thermalization) and off-diagonal (dephasing, decoherence) density-matrix
elements to their limiting values; see
 (\ref{eq2},\ref{eq3}). More generally, for large times we expect
\begin{equation}
 \rho _{nn} (t) - \rho _{nn}(\infty) \propto e^{ - t/T_{nn} } ,
\end{equation}\par\noindent
\begin{equation}
 \rho _{nm} (t) \propto e^{ - t/T_{nm} }  \qquad (n \ne m) .
\end{equation}\par\noindent
The shortest time among $T_{nn}$ is
often identified as $T_1$. Similarly, $T_2$ can be defined as the
shortest time among $T_{n \ne m}$. These definitions yield the
characteristic times of thermalization and decoherence.

For decoherence and thermalization times, the following inequality
commonly holds, $T_2  \le T_1 $,\cite{therm2} though this relation
usually does not apply within Markovian approximations. Therefore, the
decoherence time is a more crucial parameter for quantum computing
considerations. The time scale $T_2$ is compared to the ``clock''
times of quantum control, i.e., the quantum gate functions,
$T_g$, in order to ensure the fault-tolerant error correction criterion
$T_g/T_2 \leq O\left( 10^{-4}\right) $.\cite{Vincenzo}

The disadvantages of this type of
analysis are that the exponential behavior of the density matrix elements
in the energy basis is applicable only for large times, whereas
for quantum computing applications, the short-time behavior
is usually relevant.\cite{short}
Moreover, while the energy basis is natural
for large times, the choice of the preferred basis is not obvious for
short and intermediate times.\cite{short,basis} Therefore, the
time scales $T_1$ and $T_2$ have limited applicability in evaluating quantum
computing error correction criteria. Their obvious advantage is in that
of all the measures discussed in this article, they are the only experimentally
observable time scales, as long as no multi-qubit quantum computer was built.

\section{Quantum Entropy}
An alternative approach is to calculate the entropy\cite{Neumann}
of the system,
\begin{equation}
S(t)=- {\rm Tr}\left(  \rho \ln  \rho \right),
\end{equation}\par\noindent
or the idempotency defect, called also the first order entropy,\cite{Kim,Zurek,Zagur}
\begin{equation}
 \label{trace}
s(t)=1 - {\rm Tr} \left( \rho ^2 \right).
\end{equation}\par\noindent
Both expressions are basis independent, have a minimum at pure
states and effectively describe the degree of the state's
``purity.'' Any deviation from the pure state leads to the
deviation from the minimal values, 0, for both measures,
\begin{equation}
S_{\,\rm pure\ state}(t)= s_{\,\rm pure\ state}(t)= 0.
\end{equation}

Let us consider a simple
example of a two-level system in the excited state
$|1\rangle$ at $t=0$, which decays to a ground state $|0\rangle$ via
the interaction with a reservoir at zero temperature $T=0$,
e.g., the spontaneous decay of an atom in vacuum due to the
electromagnetic field. Except for very short times,\cite{therm2}
the dynamics of the system is well described by the following
density matrix,
\begin{equation}\label{atom}
 \rho _{11} (t) = e^{ - \Gamma t} , \qquad \rho _{00} (t) = 1 - e^{ - \Gamma t} ,
\end{equation}\par\noindent
\begin{equation}\label{atom1}
 \rho _{01} (t) = \rho _{10} (t) = \rho _{01} (0) = \rho _{10} (0) =
 0,
\end{equation}\par\noindent
where $\Gamma$ is the rate of the spontaneous decay.
\begin{figure}[t]
\epsfxsize=16cm\epsfbox{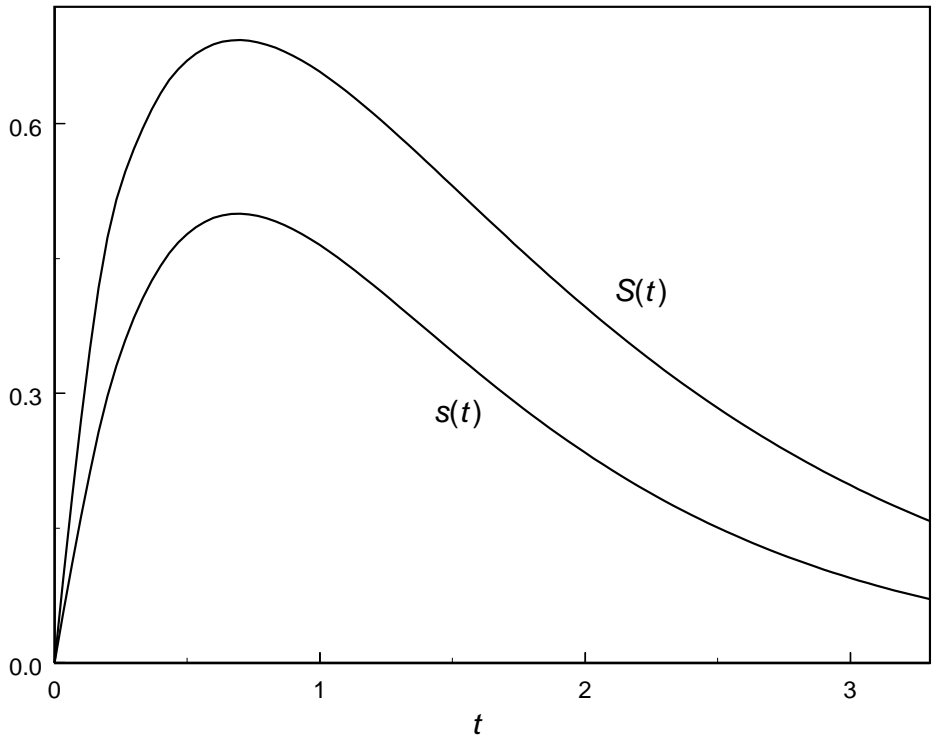}
\caption{Entropy measures of decoherence for the
two-state spontaneous decay model. The upper curve: the full entropy.
The lower curve: the first-order entropy.}
{}
\hphantom{A} 
\hrule
\end{figure}

In Figure~1, we plot the two entropy measures of
decoherence as functions of time, for $\Gamma  = 1$. Both curves
illustrate that these measures determine deviations from
``purity,'' but are not sensitive to deviations from any particular
pure (or mixed) state. It is likely that entropy measures are
useful in situations when the system has evolved little from
a particular initial pure state under its own dynamics and environmental
influences. Entropy deviations can thus be used to quantify degree of
decoherence/relaxation for short times,\cite{short} when the system
is still sufficiently far from any but the ideal pure state 
(corresponding to evolution under $H_S$ only).

One can derive\cite{Kim} the short-time perturbative expansion
for $s(t)$. Let $R(t)$ denote the density matrix of the system
plus environment, and, for definiteness, assume that the system is
initially in the pure state $|\psi \rangle$, unentangled with the
environmental bath ($B$) of modes which are described by the density
matrix $\theta$. Then,
\begin{equation}\label{e2}
\rho (t)={\rm Tr}_{\,B} R(t) ,
\end{equation}\par\noindent
\begin{equation}\label{e1}
R(t)=e^{-iHt}R(0)\, e^{iHt},
\end{equation}\par\noindent
\begin{equation}
R(0)=\left( \, |\psi \rangle \langle \psi |\, \right  ) \otimes
\theta ,
\end{equation}\par\noindent
\begin{equation}
H =   H_S  +   H_B  +   V,
\end{equation}\par\noindent
where $H$ is the total Hamiltonian, $H_S$ is the Hamiltonian of
the quantum system, $H_B$ is the Hamiltonian of the environmental
bath, $V$ is the interaction Hamiltonian. Here we use the units
$\hbar =1$, and we omit the direct product symbol in what follows. We
get the expansion of the form
\begin{equation}\label{e3}
s(t)=\left( t / \tau_s \right)^2 + o (t^2),
\end{equation}\par\noindent
\begin{equation}\label{e4}
 1/(2\tau_s^2)=
\langle \left(\langle V\rangle _B\right )^2 \rangle _S  + \langle
\left (\langle V\rangle _S \right )^2 \rangle _B  - \langle
\langle V^2 \rangle _S \rangle _B  - \left (\langle \langle
V\rangle _S \rangle _B \right )^2,
\end{equation}\par\noindent
where $\langle ...\rangle_B$, $\langle ...\rangle_S$ denote
averaging with respect to $\theta$ and $\rho(0)$, respectively.
Here, as usual, average of an operator stands for the trace of its
product with the density matrix. Obviously, $\langle \langle
...\rangle _S \rangle _B = \langle \langle
...\rangle _B \rangle _S\,$. The expression (\ref{e4}) is only
well defined when all the traces yield final results, which should
be usually the case for finite-temperature thermal initial bath
density matrix $\theta$, or for projection-operator $\theta$. More
sophisticated short-time approximations have been proposed;\cite{short}
see below.

For an instructive example, let us consider next a rather general
model of the two-level system interacting with a boson-mode
reservoir.\cite{Leggett} As usual, to evaluate the density
operator dynamics, we have to use some approximations. For short
time scales, we apply the recently developed short-time approximation,\cite{short}
rather than the straightforward perturbative
expansion (\ref{e3}). The Hamiltonian of the system has the form,
\begin{equation}\label{3}
  H =   H_S  +   H_B  +   V =-\displaystyle\frac{\Omega}{2}\sigma _z + \sum\limits_k {\omega_k
a_k^{\dagger} a_k}  +    \sigma _x \sum\limits_k {(g_k a_k^\dagger
+ g_k^* a_k )}.
\end{equation}\par\noindent
Here $  H_S={-(\Omega / 2)}\sigma _z$ is the Hamiltonian of the
quantum system, $  H_B= \sum\limits_k {\omega_k a_k^{\dagger}
a_k}$ is the Hamiltonian of the bath modes, and the remaining
term, $ V$, is the interaction Hamiltonian; $a_k,\,  a_k^\dagger$
are the boson annihilation and creation operators; $ \sigma_i\;
(i=x,y,z)$ are the Pauli matrices; $\Omega>0$ is the energy gap.
The eigenstates of $ \sigma_x$ will be denoted by
\begin{equation}\label{4}
   \sigma _x | \pm \rangle  = \pm | \pm \rangle ,
\end{equation}\par\noindent
where $|\pm\rangle= \left(|0\rangle  \pm |1 \rangle\right)/\sqrt
2$, and $|0 \rangle=\mid\uparrow \rangle,$ $|
1\rangle=\mid\downarrow \rangle$ are the ground and excited states
respectively.

Now, to obtain the short time dynamics of the system, one can use
the approximation,\cite{short}

\begin{equation}\label{short-time}
\rho _{mn} (t) = \sum\limits_{ { p, q=0,1 \atop \mu , \nu = \pm 1 }} {} \{
e^{i(E_q + E_n  - E_p  - E_m )t/2} \langle m|\mu \rangle\langle
\mu |p\rangle\langle q|\nu \rangle\langle \nu |n \rangle
\rho _{pq} (0) e^{  - B^2 (t)(\eta
_\mu - \eta _\nu  )^2 /4 + iC(t)(\eta _\mu  ^2  -
\eta _\nu  ^2 )}\},
\end{equation}\par\noindent
\begin{equation}\label{spectral function}
 B^2 (t) \equiv 8 \sum\limits_k {} \displaystyle \frac{{\left | {g_k } 
\right | ^2 }}{{\omega _k ^2 }}\sin ^2 {\displaystyle{{\omega _k t} 
\over 2}}\coth {\displaystyle{{\beta \omega _k } \over
 2}},
\end{equation}\par\noindent
\begin{equation}
 C(t) \equiv \sum\limits_k {} \displaystyle\frac{{\left| {g_k } 
\right|^2 }}{{\omega _k ^2 }}(\omega _k t - \sin \omega _k
 t).
\end{equation}\par\noindent
Here the Roman-labeled states, $|i\rangle$, are the eigenstates of 
$ H_S$ corresponding to the
 eigenvalues $E_i=(-1)^{(i+1)} \Omega /2$, with $i = m,n,p,q = 0, 1$. 
The Greek-labeled states, $|\zeta \rangle$, are the $\pm $ eigenstates 
of $\sigma_x$ in (\ref{4}), with eigenvalues $\eta _\zeta = \zeta$, where $\zeta =
\mu, \nu = \pm 1$. Evaluation of 
 (\ref{short-time}) yields the following expressions,
\begin{equation}\label{evolution}
\rho _{ 11 } (t) = \frac{{1}}{2}\left [ 1 + e^{ - B^2 (t)} \right
] \rho _{ 11 } (0) + \frac{{1}}{2}\left[1 - e^{ - B^2
(t)}\right]\rho _{ 00 } (0),
\end{equation}\par\noindent
\begin{equation}
\rho _{ 10 } (t) =
\frac{{1}}{2} e^{ -i\Omega t} \left[1 + e^{ -
B^2 (t)} \right]\rho _{ 10} (0) +
\frac{{1}}{2}\left[1 - e^{ - B^2 (t)} \right]\rho _{
01 } (0).
\end{equation}\par\noindent
It transpires that the time-dependence of the density matrix
elements within this approximation is not exponential. For
estimation of the departure from the initial pure state, we obtain
\begin{equation}\label{1qubrho}
s(t)=1-{\rm Tr}\left(\rho ^2\right)
=\displaystyle\frac{1}{2}\left[1-e^{-2B^2(t)}\right]\left\{\left[\rho
_{11}(0)-\rho _{00}(0)\right]^2+4\left| \rho _{01}(0)\right|
^2\sin ^2[(\Omega/2)t-\gamma _0]\right\},
\end{equation}\par\noindent
where $\rho_{01}=|\rho_{01}|e^{- i \gamma_0}$.

Note that this result depends on the spectral function $B^2(t)$,
defined in (\ref{spectral function}). This function is
obtained by integration over the bath mode frequencies. When the
summation in (\ref{spectral function}) is converted to
integration in the limit of infinite number of the bath modes,\cite{basis,vanKampen,Palma}
\begin{equation} \label{Bint}
 B^2 (t) = 8 \int d \omega N(\omega) |g(\omega)|^2 \omega^{-2} 
\sin ^2 {\displaystyle{{\omega t} \over 2}}
 \coth {\displaystyle{{\beta \omega} \over  2}} ,
\end{equation}\par\noindent
where $N(\omega)$ is the density of states. In many realistic
models of the bath, the density of states increases as a power of
$\omega$ for small frequencies and has a cutoff at large
frequencies (Debye cutoff in the case of a phonon bath). Therefore,
approximately setting
\begin{equation}  \label{Bint2}
 N(\omega) |g(\omega)|^2  \propto \omega^n
 \exp \left(- \omega / \omega_c \right)
\end{equation}\par\noindent
can yield a good qualitative estimate of the relaxation behavior.\cite{vanKampen,Palma}
For a popular case of Ohmic dissipation,\cite{Leggett} $n=1$, the time dependence of $B^2(t)$ is sketched
in Figure 2. One can identify the initial stage of quadratic
growth of the spectral function, the intermediate region of
logarithmic growth, and the linear large-time behavior.
\begin{figure}[t]
\epsfxsize=16cm\epsfbox{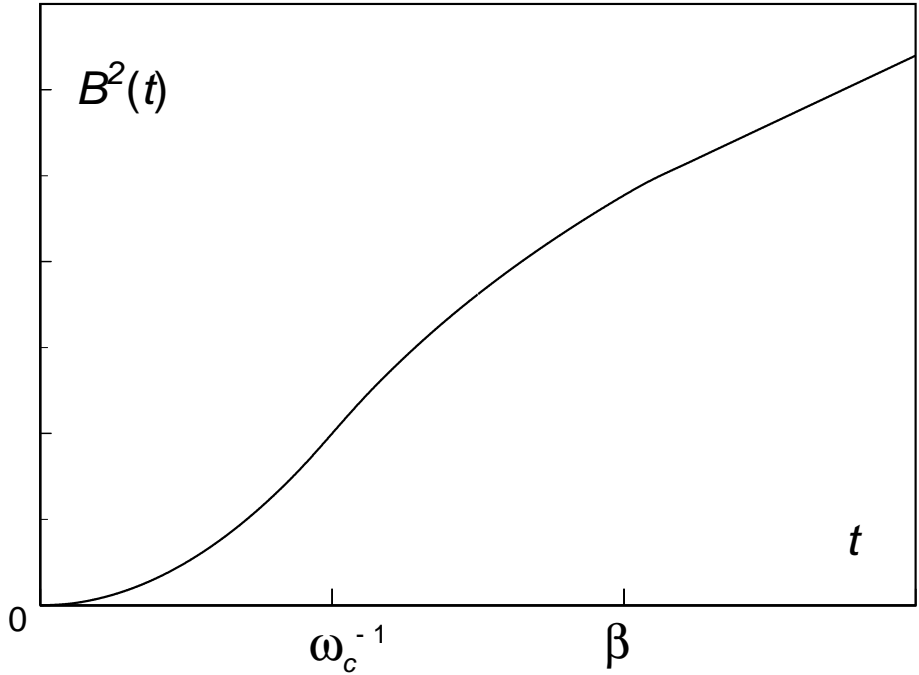}
\caption{Schematic plot of the spectral function $B^2 (t)$ for Ohmic
dissipation.\cite{Palma,basis} The ``quiet'' regime
$t < \omega^{-1}_c$ corresponds to $B^2(t)\propto (\omega_c t)^2$, the
``quantum'' regime $\omega^{-1}_c < t < \beta$ corresponds to
$B^2(t)\propto \ln(\omega_c t)$, and ``thermal'' regime $t\gg\beta,$ corresponds to $B^2(t)\propto t/\beta$.}
{}
\hphantom{A} 
\hrule
\end{figure}

The result  (\ref{1qubrho}) illustrates two common features of
the decoherence measures. The first is the explicit dependence on
the initial density matrix elements. The second is the
time-dependence that involves the spectral frequencies of the
system. We note that the spectral function $B^2(t)$ is monotonic.
However, the expression  (\ref{1qubrho}) also contains
oscillatory time dependence with the period $2\pi/\Omega$, which
obviously reflects the property of the system's internal dynamics
rather than its deviation from a pure state. If the initial
density matrix is diagonal, $\rho_{01}=0$, then we have $s(t)=
2^{-1}\{1-[\rho _{00 } (0) - \rho _{ 11 } (0)]^2 e^{ - 2B^2
(t)}\}$, which has no such oscillations.

\section{Fidelity}
\label{Fidelity}

Writing the total Hamiltonian as usual,
\begin{equation}\label{f0}
  H=H_S+H_B+V,
\end{equation}\par\noindent
let us now define the fidelity,\cite{Fidelity,Fidelity2}
\begin{equation}\label{f1}
F(t)={\rm Tr}_{\,S}  \left[ \, \rho _{\rm ideal}(t) \, \rho (t) \, \right],
\end{equation}\par\noindent
where the trace is over the system degrees of freedom, and $\rho_{\rm ideal}(t)$
represents the pure-state evolution of the system under the constant $H_S$ only,
without interaction with the environment ($V=0$),
\begin{equation}\label{f2}
\rho _{\rm ideal}(t)= e^{-iH_S t}\rho(0)\, e^{iH_S t}.
\end{equation}

The fidelity provides a certain measure of decoherence in terms of
the difference between the ``real,'' environmentally influenced,
$\rho (t)$, evolution and the ``free'' evolution, $\rho_{\rm
ideal} (t)$. It will attain its maximal value, 1, only provided
$\rho (t) = \rho_{\rm ideal} (t)$. This property relies on the
fact the $ \rho_{\rm ideal} (t)$ remains a projection operator
(pure state) for all times $t \geq 0$.

Let us consider the two-level system decaying from the excited to
ground state,  (\ref{atom},\ref{atom1}). In this case, there
is no internal system dynamics,
\begin{equation}
\rho _{\rm ideal} (t) =\left(
\begin{array}{cc}
0 & 0 \\
0 & 1
\end{array}
\right),
\end{equation}\par\noindent
\begin{equation}
\rho (t)=\left(
\begin{array}{cc}
1-e^{-\Gamma t} & 0 \\
0 & e^{-\Gamma t}
\end{array}
\right),
\end{equation}\par\noindent
and the fidelity is a monotonic function of time,
\begin{equation}\label{f3}
F(t)=e^{-\Gamma t}.
\end{equation}\par\noindent

Note that the requirement that $\rho_{\rm ideal}(t)$ is pure-state (projection operator), 
excludes any $T>0$ thermalized state as the initial system state. 
For example, let us consider the infinite-temperature initial
state of our two level system. We have
\begin{equation} \rho (0)=\rho _{\rm ideal}(t)=\left(
\begin{array}{cc}
1/2 & 0 \\
0 & 1/2
\end{array}
\right),
\end{equation}\par\noindent
which is not a projection operator. The spontaneuos-decay density matrix is then
\begin{equation}
\rho (t)=\left(
\begin{array}{cc}
1-(e^{-\Gamma t}/2) & 0 \\
0 & e^{-\Gamma t}/2
\end{array}\right).
\end{equation}\par\noindent
The fidelity is constant,
\begin{equation}\label{f3.1}
F(t)=1/2,
\end{equation}\par\noindent
and it does not provide any useful information of the time-dependence of the decay process.

We can derive the short-time perturbative expansion\cite{Fidelity}
 for $F(t)$,
\begin{equation}
F(t)=1-(t / \tau _F )^2+o (t^2),
\end{equation}\par\noindent
\begin{equation}\label{f5}
1/\left(2\tau _F^2\right)=\langle\langle V^2\rangle _S\rangle
_B-\langle\left(\langle V\rangle_S\right )^2\rangle_B,
\end{equation}\par\noindent
which should be compared to (16).

Let us now turn to the example of the two-level system in the
short-time approximation; see  (\ref{3}), et seq. Coherent
evolution of the two-level system due to the Hamiltonian $H_S$, is
described by
\begin{equation}\label{free}
\rho_{\rm ideal}(t)=\left(
\begin{array}{cc}
\rho _{00}(0) & \rho _{01}(0)e^{i\Omega t} \\
\rho _{10}(0)e^{-i\Omega t} & \rho _{11}(0)
\end{array}
\right).
\end{equation}\par\noindent
A straightforward calculation gives
\begin{equation}\label{f6}
F(t)=1 - \displaystyle\frac{1}{2}\left[1-e^{-B^2(t)}\right]\left\{\left[\rho
_{11}(0)-\rho _{00}(0)\right]^2+4\left| \rho _{01}(0)\right|
^2\sin ^2[(\Omega/2)t-\gamma _0]\right\}.
\end{equation}\par\noindent
This result should be compared to (\ref{1qubrho}) While not
identical, it has similar features, including the system-frequency
oscillations.

\section{Norm of deviation}
In this section we propose to use the operator norms \cite{Kato}
that measure the deviation of the system from the ideal state, to
quantify the degree of decoherence. Such measures do not require the
initial density matrix to be pure-state. We define the deviation
according to
\begin{equation}\label{deviation}
  \sigma(t)  \equiv   \rho(t)  -   \rho_{\rm ideal} (t)  .
\end{equation}\par\noindent
We can use, for instance, the
eigenvalue norm,
\begin{equation}\label{n11}
\left\|\sigma \right\|_{\lambda} = {\max_i} \left|
{\lambda _i } \right|,
\end{equation}\par\noindent
or the trace norm,
\begin{equation}\label{tracenorm}
\left\|   \sigma  \right\|_{{\rm Tr}}  = \sum\limits_i {\left|
{\lambda _i } \right|},
\end{equation}\par\noindent
etc., where $\lambda_i$ are the eigenvalues of the deviation
operator,  (\ref{deviation}).

For more precise definitions, let us consider an arbitrary linear
operator $A$. One of the possible ways to define the norm of $A$ is\cite{Kato}
\begin{equation}\label{n2}
\left\| A \right\| = \mathop {\sup }\limits_{\varphi  \ne 0}
\left[ {\frac{{ \langle \varphi |A^ \dagger  A|\varphi  \rangle }}{{
\langle \varphi |\varphi  \rangle }}} \right]^{1/2}.
\end{equation}\par\noindent
Since density operators are bounded, their norms, as well
the norm of the deviation, can be always evaluated. Furthermore,
since the density operators are Hermitean, this definition
obviously reduces to the eigenvalue norm (41). We also note that
$\left\| A \right\|=0$ implies that $A=0$.

The calculation of these norms is sometimes simplified by the observation that
$\sigma(t)$ is traceless. Specifically, for two-level systems, we get
\begin{equation} \left\|   \sigma
\right\|_{\lambda} = \sqrt {\left| {\sigma _{00} } \right|^2
+ \left| {\sigma_{01} } \right|^2 } = {1 \over 2}
\left\|   \sigma  \right\|_{{\rm Tr}}.
\end{equation}\par\noindent
Therefore, for tow-state systems, we will only consider the eigenvalue norm.
For our example of the two-level system undergoing spontaneous decay,
the norm is
\begin{equation}
\left\|   \sigma
\right\|_{\lambda} = 1 - e^{-\Gamma t} .
\end{equation}\par\noindent
Thus, in this case $\left\|   \sigma
\right\|_{\lambda} = 1 - F(t)$.

Next, consider the two-level system coupled to a bath of modes, in
the short-time approximation; see  (\ref{3}), et seq.
The coherent (ideal) evolution of the system due to the
Hamiltonian $H_S$, is described by  (\ref{free}). One can then
obtain the result
\begin{equation}\label{n1qubit}
\left\| \sigma(t)  \right\|_{\lambda} = \frac{1}{2}\left[ 1 - e^{
- B^2 (t)}\right] \left\{\left[\rho _{11} (0) - \rho _{00}
(0)\right]^2 + 4 \left| \rho _{01} (0) \right|^2 \sin ^2
[(\Omega /2)t -\gamma_0]\right\}^{1/2},
\end{equation}\par\noindent
which should be compared to  (\ref{1qubrho}) and (\ref{f6}). At
$t=0$, the value of the norm is equal to 0, and then it increases
to positive values, with superimposed modulation at the system's
energy-gap frequency.

\section{Arbitrary Initial States, Multiqubit Systems}

The measures considered in the preceding sections quantify decoherence of a system
provided its initial state is given. However, this is not always the
case. Usually, it will be necessary to obtain an upper-bound estimate of
decoherence for an arbitrary initial state. For example, in
quantum teleportation schemes, the quantum state to be teleported is
not known. In quantum computing, the ideal state of
a quantum register is theoretically specified for each
step of a quantum algorithm. But from the practical point of view,
the bookkeeping becomes intractable even for few-gate algorithms.
Furthermore, even the preparation of the initial state can introduce noise.

To characterize decoherence for an arbitrary initial state, pure or mixed, we propose to use
the maximal norm, $D$, which is determined as an operator
norm maximized over all initial density matrices. For instance, we can define
\begin{equation}\label{normD}
  D(t) = \sup_{\rho (0)}\bigg(\left\| \sigma (t,\rho (0))\right\|_{\lambda}  \bigg).
\end{equation}\par\noindent
For the two-level system coupled to a bosonic bath, with the dynamics 
described by the Hamiltonian (\ref{3}), the expression of the maximal
norm is elegant and compact,
\begin{equation}\label{D1qubit}
D(t)  = \frac{1}{2}\left[ 1 - e^{ - B^2 (t)}\right] .
\end{equation}\par\noindent
The result is monotonic and contains no oscillations due to the
internal system dynamics, as shown in Figure 3.
\begin{figure}[t]
\epsfxsize=16cm\epsfbox{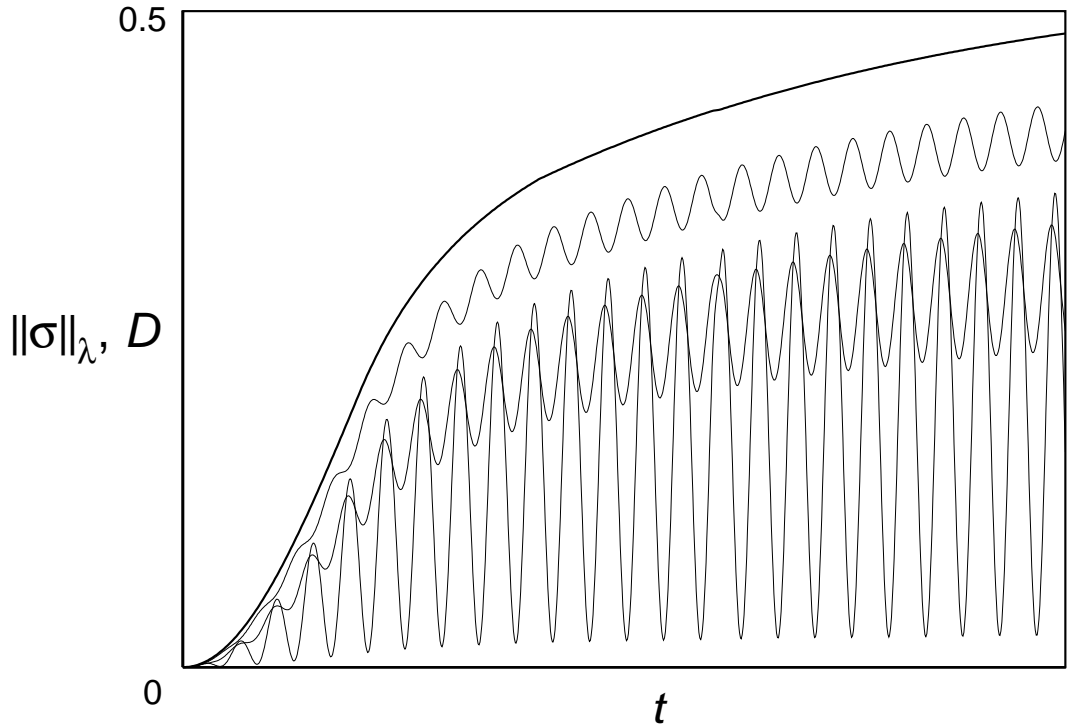}
\caption{Norm of the deviation from the ideal dynamics for the density
matrix of a spin interacting with an Ohmic bath of bosonic modes, in the
short-time approximation. The upper curve: the maximal norm $D(t)$; 
the lower curves: the $\left\| \sigma
\right\|_{\lambda}$ norms for three different initial states $\rho
(0)$, chosen to illustrate the overall pattern.}
{}
\hphantom{A} 
\hrule
\end{figure}

In principle, the calculation of the maximal norm for a large-scale
multiqubit system should be formidable task, since it implies
maximization over an exponentially large (in the number of
qubits) set of coefficients of the initial density matrix.
However, for specific applications, we can use approximate upper bounds
calculated by evaluating norms for small subsystems.
Consider a multiqubit system $S$ consisting of two separate
{\em unentangled\/} subsystems $S_2$ and $S_2$, at time $t$, with decoherence norms
$D_1$ and $D_2$, respectively. Let us denote the density matrix of the
full system and its deviation as $\rho$ and $\sigma$, respectively, and
use the same notation with indices 1 and 2 for the two subsystems. For
brevity, we use the superscript $(i)$ to denote the ``ideal'' density 
matrices. The overall norm $D(t)$ can then be written as
\begin{equation}\label{D12}
  D = \sup_{\rho (0)} \left(\left\| \sigma \right\|_{\lambda}
  \right)
  = \sup_{\rho (0)} \left(\left\| \rho - \rho ^{(i)}\right\|_{\lambda}
  \right)
  = \sup_{\rho (0)} \left(\left\| \rho_1   \rho_2 - \rho_1 ^{(i)}   \rho_2 ^{(i)}\right\|_{\lambda}
  \right)
  = \sup_{\rho (0)} \left(\left\| \sigma_1   \rho_2 + \rho_1 ^{(i)}   \sigma_2\right\|_{\lambda}  \right)
\end{equation}\par\noindent
and estimated by using the following sequence of inequalities,
\begin{equation}\label{ineq1}
  D= \sup_{\rho (0)} \left(\left\| \sigma_1   \rho_2 + \rho_1 ^{(i)}   \sigma_2\right\|_{\lambda}
  \right)
  \leq
  \sup_{\rho (0)} \left(\left\| \sigma_1   \rho_2 \right\|_{\lambda}
  \right)
  +
  \sup_{\rho (0)} \left(\left\|  \rho_1 ^{(i)}   \sigma_2\right\|_{\lambda}  \right),
\end{equation}\par\noindent

\begin{equation}\label{ineq2}
\sup_{\rho (0)} \left(\left\| \sigma_1   \rho_2
\right\|_{\lambda}
  \right)
  +
  \sup_{\rho (0)} \left(\left\|  \rho_1 ^{(i)}   \sigma_2\right\|_{\lambda}  \right)
  \leq
\sup_{\rho_1 (0)}\left(\left\| \sigma_1
\right\|_{\lambda}
  \right)
  +
  \sup_{\rho_2 (0)}\left(\left\| \sigma_2\right\|_{\lambda}
  \right)=
  D_1 + D_2 \, .
\end{equation}\par\noindent
The inequality $D \leq D_1 + D_2$ is, of course, approximate, because we have assumed that the
interaction with the bath, and with each other, left the subsystems unentangled at
time $t$. However, it provides an indication that, perhaps with
a smart choice of a decomposition into subsystems which are initially not significantly
entangled, the maximal norm can be approximately considered an extensive quantity, linear in the number of
qubits, at least for short times. It is expected that for larger times, the relaxation process {\em rates\/}
are the additive quantities growing linearly with the number of qubits, as long as no quantum
error correction is involved.

In conclusion, we have considered several approaches to
quantify decoherence: relaxation times, entropy and fidelity measures, 
and norms of deviation. The latter measures offer certain
advantages, and we defined the maximal measure that is not
dependent on the initial state and, at least for short times,
is approximately extensive in the number of qubits.

This research was supported by the National Security Agency and
Advanced Research and Development Activity under Army Research
Office contract DAAD-19-02-1-0035, and by the National Science
Foundation, grants DMR-0121146 and ECS-0102500.

\end{document}